\documentclass[12pt]{article}
\usepackage[utf8]{inputenc}
\usepackage[russian,english]{babel} 
\usepackage[T2A]{fontenc}  
\usepackage{amsfonts,amssymb,amsmath} 
\usepackage{graphicx} 
\usepackage{hyperref} 
\hypersetup{hidelinks} 
\usepackage{mathtools}
\usepackage{cite}  

\allowdisplaybreaks 
\makeatletter \@addtoreset{equation}{section}
\renewcommand\theequation
  {\ifnum \c@section>\z@ \thesection.\fi \@arabic\c@equation}
\renewcommand{\@biblabel}[1]{#1.}
\makeatother

\begin{document}

\title{\setcounter{footnote}{0}\Large\bf Gradient Projection Method 
and Stochastic Search \\
for Some Optimal Control Models \\
with Spin Chains.~II\footnote{This research was funded by the Russian Science Foundation, project No.~22-11-00330-P, and performed at the Steklov Mathematical Institute of Russian Academy of Sciences.}}

\author{\setcounter{footnote}{6}\bf Oleg~V.~Morzhin\footnote{\footnotesize E-mail: \url{morzhin.oleg@yandex.ru};~ 
   \href{http://www.mathnet.ru/eng/person30382}{mathnet.ru/eng/person30382};~ 
    \href{https://orcid.org/0000-0002-9890-1303}{ORCID 0000-0002-9890-1303}} 
    \vspace{0.2cm} \\
\normalsize Department of Mathematical Methods for Quantum Technologies, \vspace{-0.1cm} \\ 
\normalsize Steklov Mathematical Institute of Russian Academy of Sciences, \vspace{-0.1cm} \\
\normalsize Gubkina Str.~8, Moscow, Russian Federation} 

\date{}
\maketitle
\vspace{-0.8cm}

\begin{abstract}
\small This article (II) continues the research described in [Morzhin~O.V. Gradient projection method and stochastic search for some optimal control models with spin chains.~I (submitted)] (Article~I), derives the needed finite-dimensional gradients corresponding to the infinite-dimensional gradients obtained in Article~I, both for transfer and keeping problems at a~certain $N$-dimensional spin chain, and correspondingly adapts a~projection-type condition for optimality, gradient projection method (GPM). For the case $N=3$, the given in this article examples together with Example~3 in Article~I show that: a)~the adapted GPM and genetic algorithm (GA) successfully solved numerically the considered transfer and keeping problems; b)~the two- and three-step GPM forms significantly surpass the one-step GPM. Moreover, GA and a~special class of controls were successfully used in such the transfer problem that $N=20$ and the final time is not assigned.
\par Данная статья (II) продолжает исследование, описанное в [Моржин~О.В. Метод проекции градиента и стохастический поиск для некоторых моделей оптимального управления со спиновыми цепочками.~I (направлена)] (статья~I), выводит необходимые конечномерные градиенты, соответствующие бесконечномерным градиентам, полученным в статье~I для задач переноса и удержания на определенной $N$-мерной спиновой цепочке, и соответственно адаптирует некоторое условие оптимальности проекционного типа, метод проекции градиента (МПГ). Для случая $N=3$ данные в этой статье примеры вместе с примером~3 из статьи~I показывают, что: а)~адаптированный МПГ и генетический алгоритм (ГА) успешно  численно решили рассмотренные задачи переноса и удержания; б)~двух- и трехшаговые формы МПГ значительно превосходят одношаговый МПГ. Более того, ГА и специальный класс управлений были успешно использованы в такой задаче переноса, в которой $N=20$ и финальный момент не~задан.
\vspace{0.15cm}\par {\bf Keywords:} quantum optimal control, gradient projection method, genetic algorithm, spin chains.
\par {\bf Ключевые слова:} квантовое оптимальное управление, метод проекции градиента, генетический алгоритм, спиновые цепочки. 
\end{abstract}
\normalsize

\rightline{\it Dedicated to the memory of Prof. V.A.~Dykhta (1949--2025)~\cite{Morzhin_AiT_2025},}
\rightline{\it whose contribution to the mathematical theory of optimal control and its} 
\rightline{\it applications is impressive, relates, in particular,  to quantum control} 

\vspace{-0.1cm} 
\section{Introduction-II}  
\vspace{-0.4cm}

~\par Quantum optimal control (QOC) is an~important scientific direction significantly using the mathematical theory of optimal control (MTOC),~etc.  Introduction in Article~I~\cite{Morzhin_ArticleI} notes a~number of various optimization methods with a~number of references. Among various optimization tools for various quantum optimization problems, note GPM in its first-order form developed in the pioneering article~\cite{Oza_Pechen_Dominy_et_al_JPA_2009} concerning the representation of coherent and incoherent controls by complex Stiefel manifolds. Also note one-step form (GPM-1S)~\cite{Morzhin_Pechen_LJM_2019, Goncalves_GomesRuggiero_Lavor_2016, Pechen_LJM_2025}, two-step form (GPM-2S)~\cite{Morzhin_Pechen_JPA_2025, Morzhin_Pechen_QIP_2023, Bolduc_Knee_et_al_2017}, and three-step form (GPM-3S)~\cite{Morzhin_Pechen_JPA_2025, Morzhin_ArticleI}, as well as genetic algorithm~(GA)~\cite{Pechen_Rabitz_2006, Pawela_Sadowski_QIP_2016,  Morzhin_ArticleI}. 

The notion ``spin chain'' is important in the quantum science. The advanced description~\cite{Nobel_Prize_in_Physics_2016} for the Nobel Prize in Physics 2016 is, in particular, about spin chains. As \cite{Murphy_Montangero_Giovannetti_Calarco_2010} (M.~Murphy, S.~Montangero, V.~Giovannetti, T.~Calarco, 2010) notes, using spin chains as quantum channels for communication between two parties was first proposed in \cite{Bose_2003} (S.~Bose, 2003), after which the author can list a~good number of publications in the direction for transfer along various spin chains, e.g., 
\cite{Bose_2007, Gong_Brumer_2007, Caneva_Murphy_Calarco_et_al_2009, Murphy_Montangero_Giovannetti_Calarco_2010, Chakrabarti_VanderJeugt_2010, Trushkova_AiT_2013, Gurman_Rasina_Baturina_IFAC_2013, Gurman_Guseva_Fesko_AIPConfProc_2016,  Feldman_Pechen_Zenchuk_2021, Feldman_Pechen_Zenchuk_2024,  Stepanenko_Chernova_Gorlach_PRL_2025}. 

This article continues to analyze the transfer and keeping problems given in Sect.~2 in Article~I and based on \cite{Caneva_Murphy_Calarco_et_al_2009, Murphy_Montangero_Giovannetti_Calarco_2010, Trushkova_AiT_2013}. These problems contains the Cauchy problem (Article~I,~2.1), i.e. \vspace{-0.2cm}
\begin{equation*}
\frac{d\psi^u(t)}{dt} = -i \big(H_0 + H_1(t,u(t))\big)\psi^u(t), \quad \psi^u(0)=\psi_0, \vspace{-0.1cm}
\end{equation*}
for such the $N$-level Schr\"{o}dinger equation that its full Hamiltonian $H$ contains the part $H_1$ depending on control $u=(u_1, u_2)$ and explicitly on~$t$. 

An~important aspect is how to solve this Cauchy problem at a~given~$u$ and solve the given in [Lemma~1, Article~I] adjoint dynamic equations. For some another quantum optimal control problems (OCPs), \cite{Morzhin_Pechen_LJM_2019, Morzhin_Pechen_QIP_2023,  Morzhin_Pechen_JPA_2025}, etc. use such the universal direction that, first, either realificates a~considered quantum OCP or realificates those GPM constructions which are for the initial OCP and, second, use some standard numerical method (e.g., the 5(4) Runge--Kutta method) for solving the corresponding Cauchy problems with the real states. This approach takes into account that it can be a~Hamiltonian explicitly depending on time, piecewise linear interpolation for controls, etc. The formula (Article~I,~4.1) for $\psi^u(t_j)$ in the matrix exponentials' terms under piecewise constant (PConst.) controls is without any realification, reminds the corresponding stage of GRAPE in \cite[Eq.~(29)]{Khaneja_et_al_2005} (N.~Khaneja, T.~Reiss,  C.~Kehlet, et al., 2005), Geodesic Pulse
Engineering in \cite{Lewis_Wiersema_Bose_2025}~(D.~Lewis, R.~Wiersema, S.~Bose, 2025), and GPM-1S in \cite{Pechen_LJM_2025}~(A.N.~Pechen, 2025) for some another OCPs. In contrast, the constructed below final-dimensional (fin.-dim.) GPM needs to solve the adjoint systems (Article~I,~3.1),~(Article~I,~3.2) under PConst. controls where the way being by analogy with~(Article~I,~4.1) is used. For the infinite-dimensional (infin.-dim.) gradients (Article~I,~3.3), this approach allows below (Lemma~1) to derive the corresponding fin.-dim. gradients, both for the transfer and keeping problems. Thus, an important aspect in Article~II is to realize the idea known in MTOC (e.g., \cite{Buldaev_2021} (A.S.~Buldaev, 2021), \cite{Arguchintsev_Srochko_2022}\,(A.V.~Arguchintsev, V.A.~Srochko, 2022)) and meaning that, under PConst. controls, a~fin.-dim. gradient can be obtained from the infin.-dim. gradient derived for the corresponding OCP under piecewise continuous controls. The given below in Lemma~1 gradients are used below in the corresponding fin.-dim. GPM forms. Moreover, this article uses the special class~(Article~I,~2.11), the given in~(Article~I,~2.12) objective function~$f_4$, and the corresponding GA~approach. This article does not analyze thoroughly the iterative GPM behavior, but below Examples~1,\,2 show that here the adapted GPM forms approximately solve the transfer and keeping problems under $N=3$ and GPM-2S, GPM-3S are significantly faster than GPM-1S. 

This article starts numbering again, use the references ``(Article~I,~2.1)'', etc., does not repeat the given in [Article~I, Sect.~2] statements for the transfer and keeping problems. The structure of Article~II is as follows. Sect.~\ref{Sect2} gives the fin.-dim. gradients, GPM, some condition for optimality. Sect.~\ref{Sect3} represents the interesting GPM and GA results obtained for the case $N=3$. Sect.~\ref{Sect4} considers the case $N=20$ and the GA successful results. Sect.~\ref{Sect5} gives some important methodological comments.

\vspace{-0.1cm}
\section{Finite-Dimensional Gradients, Projection-Type Condition for Optimality, and GPM Forms}
\label{Sect2}
\vspace{-0.4cm}

~\par{\bf Lemma 1.} 
\label{lemma1}
{\it Consider the system (Article~I,~2.1)--(Article~I,~2.3) with PConst. $u$, the certain $\sigma$ at a~given time grid~$\Theta_M$ together with the given PConst. $b_l, S_l$, $l=1,2$ in (Article~I,~2.4), (Article~I,~2.9). Both for the transfer and keeping problems in the terms of the objective functions $f_p({\bf a})$, $p =1,2$, we have: 1)~the increment formula (Article~I,~3.4) in the form \vspace{-0.2cm}
\begin{equation}
\hspace{-0.0cm} \Phi_p(u) - \Phi_p(u^{(k)}) = f_p({\bf a}) - f_p({\bf a}^{(k)}) = \big\langle {\rm grad}\, f_p({\bf a}^{(k)}), {\bf a} - {\bf a}^{(k)} \big\rangle + r,
\hspace{-0.2cm} 
\label{sect7_f1} 
\vspace{-0.2cm}
\end{equation}
where the $s$th gradient component is continuous in ${\bf a}$ and has the form \vspace{-0.2cm}
\begin{equation}
({\rm grad}\, f_p({\bf a}^{(k)}))_s = \int\nolimits_{t_j}^{t_{j+1}} G_l^{(k)}(t;{\bf a})dt, ~~G_l^{(k)} = ({\rm grad}\, \Phi_p(u^{(k)})(t))_l, \vspace{-0.2cm}
\label{sect7_f2}
\end{equation}
the indices $p \in \{1,2\}$, $s \in \overline{1, 2M}$, $j \in \overline{0,M-1}$, $l \in \{1,2\}$; 
2)~for a~segment $[t_j, t_{j+1}]$, $j \in \overline{0,M-1}$, the value of $G_l^{(k)}$ at $t = (t_j+t_{j+1})/2$ \vspace{-0.15cm}
\vspace{-0.2cm}
\begin{align}
&G_l^{(k)}\Big(\frac{t_j+t_{j+1}}{2};{\bf a}\Big) = -{\rm Im} \Big\langle \eta^{c^{(k)}_{j+1}}\Big(\frac{t_j+t_{j+1}}{2}\Big), \frac{\partial}{\partial u_l}H_1(t,u)\big|_{u=c^{(k)}_{j+1}} \times \nonumber \\ 
& \times \psi^{c^{(k)}_{j+1}}\Big(\frac{t_j+t_{j+1}}{2}\Big) \Big\rangle_{\mathbb{C}^N} + 2 P_{u_l} S_l \Big(\frac{t_j+t_{j+1}}{2}\Big) c^{(k)}_{l,j+1},  
\label{sect7_f3}\\
\text{where} &~\psi^{c^{(k)}_{j+1}}((t_j + t_{j+1})/2) = e^{A(c^{(k)}_{j+1})\,((t_{j+1} - t_j)/2)}\, \psi^{c^{(k)}_j}(t_j) 
\label{sect7_f4} 
\end{align}
and, for the transfer problem, take \vspace{-0.2cm}
\begin{align}
&\eta^{c^{(k)}_{j+1}}((t_j + t_{j+1})/2) = e^{-A(c^{(k)}_{j+1})\,((t_{j+1} - t_j)/2)}\, \eta^{c^{(k)}_{j+2}}(t_{j+1})
\label{sect7_f5}\\
&\text{with}~~\eta^{c^{(k)}_{j+2}}(t_{j+1}) = e^{-A(c^{(k)}_{j+2})(t_{j+2} - t_{j+1})} \nonumber \times \\
& \times \eta^{c^{(k)}_{j+3}}(t_{j+2}), ~j \in \overline{0,M-3}, ~~ \eta^{c^{(k)}_M}(t_M) = \eta^{u^{(k)}}_{\rm Transv.},
\label{sect7_f6}
\end{align}
while, for the keeping problem, take \vspace{-0.2cm}
\begin{align}
&\eta^{c^{(k)}_{j+1}}((t_j+t_{j+1})/2) = e^{-A(c^{(k)}_{j+1})(t_{j+1}-t_j)/2} \eta^{c^{(k)}_{j+1}}(t_{j+1}) - \nonumber\\
&\quad- P_{\psi}\int\nolimits_{(t_j+t_{j+1})/2}^{t_{j+1}} e^{A(c^{(k)}_{j+1})((t_j+t_{j+1})/2 - \zeta)} \langle \psi_{\rm g}, \psi^{c^{(k)}_{j+1}}(\zeta) \rangle \psi_{\rm g}\,d\zeta.
\label{sect7_f7}
\end{align}}

{\bf Proof.} 1)~For (\ref{sect7_f1}),~(\ref{sect7_f2}) under admissible PConst. $u,\,u^{(k)}$, one has \vspace{-0.2cm}
\begin{eqnarray*}
&&f_p({\bf a}) - f_p({\bf a}^{(k)}) =\Phi_p(u) - \Phi_p(u^{(k)}) = \\
&&= \sum\nolimits_{j=0}^{M-1} \sum\nolimits_{l=1}^{2} \Big( \int\nolimits_{t_j}^{t_{j+1}} G_l^{(k)}(t;{\bf a}) dt \Big)_l \big(u_l(t_j) - u_l^{(k)}(t_j) \big) + r =\\
&&= \sum\nolimits_{s=1}^{2M} ({\rm grad}\, f_p({\bf a}^{(k)}))_s \, ({\bf a}_s - {\bf a}_s^{(k)}) + r = \big\langle {\rm grad}\, f_p({\bf a}^{(k)}), \, {\bf a} - {\bf a}^{(k)} \big\rangle + r. \vspace{-0.2cm}
\end{eqnarray*}
The continuity of $({\rm grad}\, f_p({\bf a}^{(k)}))_s$ in ${\bf a}$ is provided by the continuous dependence $G_l^{(k)}(t;{\bf a})$ on~${\bf a}$. 2)~(\ref{sect7_f3})--(\ref{sect7_f7}) are derived basing on (Article~I,~3.3). Here we take $c^{(k)}_{j+1}$ during a half of $[t_j, t_{j+1}]$. The formula (\ref{sect7_f5}) is constructed with continuous connecting the neighboring parts of $\eta^{u^{(k)}}$ and, in contrast to (Article~I,~4.1), realizes the movements along the time grid from $t_M$ to left, starting from $\eta^{u^{(k)}}_{\rm Transv.}$. The formula (\ref{sect7_f7}) takes into account that the adjoint equation in (Article~I,~3.1) for the keeping problem is inhomogeneous.~$\square$

The next theorem is based on the derived gradients~(\ref{sect7_f2}) ($p \in \{1,2\}$) and the projection-type 1st-order necessary condition for optimality known (e.g., \cite[part\,2, p.\,99]{ZhadanVG_3books_2024}) in the theory of fin.-dim. optimization and does not require convexity/ strong convexity for $f_p({\bf a})$ at the whole~$Q_{\bf a}$. 

\vspace{-0.2cm}

~\par{\bf Theorem 1.}  
\label{theorem1}
(Necessary 1st order condition for local optimality). 
{\it For the transfer or  keeping problem considered in Lemma~1 in the terms of $f_p({\bf a})$, $p \in \{1,2\}$, consider the provided by Lemma~1 gradient at a~given point ${\bf a}_{\ast} \in Q_{\bf a}$:  ${\rm grad}\,f_p({\bf a}_{\ast})$. Under this gradient, if ${\bf a}_{\ast}$ is a~local minimum point of $f_p({\bf a})$, then the following equality is hold: \vspace{-0.2cm}
\begin{equation}
{\bf a}_{\ast} = {\rm Pr}_{Q_{\bf a}}\big({\bf a}_{\ast} - \alpha\, {\rm grad}\, f_p({\bf a}_{\ast}) \big), ~~ \alpha > 0. 
\label{sect7_f8}  
\end{equation} }

\vspace{0.2cm}

In view of (Article~I,~3.6)--(Article~I,~3.8), the obtained fin.-dim. gradients, and (\ref{sect7_f8}), take \vspace{-0.2cm}
\begin{align}
&\text{GPM-1S:~~~}{\bf a}^{(k+1)} = {\rm Pr}_{Q_{\bf a}}\big({\bf a}^{(k)} - \alpha_k \, {\rm grad}\, f_p({\bf a}^{(k)}) \big), ~~ k \geq 0, \label{sect7_f9}  \\
&\text{GPM-2S:~~~}{\bf a}^{(k+1)} = {\rm Pr}_{Q_{\bf a}}\big({\bf a}^{(k)} - \alpha_k \, {\rm grad}\, f_p({\bf a}^{(k)}) + \nonumber \\
&\qquad\qquad\qquad\qquad\qquad + \beta_k ({\bf a}^{(k)} - {\bf a}^{(k-1)})\big), ~~k \geq 1, \label{sect7_f10} \\
&\text{GPM-3S:~~~}{\bf a}^{(k+1)} = {\rm Pr}_{Q_{\bf a}}\big({\bf a}^{(k)} - \alpha_k \, {\rm grad}\, f_p({\bf a}^{(k)}) + \beta_k ({\bf a}^{(k)} - {\bf a}^{(k-1)}) + \nonumber \\
&\qquad\qquad\qquad\qquad\qquad + \gamma_k ({\bf a}^{(k-1)} - {\bf a}^{(k-2)})\big), ~~k \geq 2. \label{sect7_f11}
\end{align} 
It is of interest, e.g., to try GPM-2S with its fixed $\alpha_k=\alpha, ~\beta_k=\beta$.

Each component $({\rm grad}\, f_p({\bf a}^{(k)}))_s$ determined in~(\ref{sect7_f2}) is approximately computed here via the midpoint rectangle method: \vspace{-0.2cm}
\begin{equation}
\hspace{-0.0cm}
\int\nolimits_{t_j}^{t_{j+1}} G_l^{(k)}(t;{\bf a})dt \approx {\rm ApprGr}(c^{(k)}_{l,j+1}) = (t_{j+1} - t_j) G_l^{(k)}\Big(\frac{t_j+t_{j+1}}{2};{\bf a}\Big), 
\hspace{-0.2cm}
\vspace{-0.2cm} 
\label{sect7_f12}
\end{equation}
where if the keeping problem is analyzed, then the integral used in (\ref{sect7_f7}) is approximated here via the trapezoidal method, i.e. as \vspace{-0.2cm}
\begin{align}
&P_{\psi}\frac{t_{j+1} - t_j}{2}\Big[\Big\langle \psi_{\rm g}, \psi^{c^{(k)}_{j+1}}\Big(\frac{t_j+t_{j+1}}{2}\Big) \Big\rangle \psi_{\rm g} + \nonumber \\
&\quad + e^{-A(c^{(k)}_{j+1})((t_{j+1}-t_j)/2)} \, \langle \psi_{\rm g}, \psi^{c^{(k)}_{j+1}}(t_{j+1}) \rangle \psi_{\rm g} \Big] = \nonumber\\
&= P_{\psi}\frac{t_{j+1} - t_j}{2}\Big[ \big( 0, ..., 0, \big(\psi_N^{c_{j+1}^{(k)}}\big(\frac{t_j+t_{j+1}}{2}\big)\big)^{\ast}\big)^{\rm T} + \nonumber\\
&\quad + e^{-A(c^{(k)}_{j+1})((t_{j+1}-t_j)/2)} \, \big(0, ..., 0, \big(\psi_N^{c_{j+1}^{(k)}}(t_{j+1})\big)^{\ast}\big)^{\rm T} \Big].
\label{sect7_f13}
\end{align}
Thus, for the schemes (\ref{sect7_f9})--(\ref{sect7_f11}), consider the following schemes using (\ref{sect7_f12}), (\ref{sect7_f13}) (here $l =1,2$, $j = \overline{1,M-1}$):  \vspace{-0.25cm}
\begin{align*}
&c^{(k+1)}_{l,j+1}={\rm Pr}_{[-\nu_{l,j}, \nu_{l,j}]}\big(c^{(k)}_{l,j+1} - \alpha_k \, {\rm ApprGr}(c^{(k)}_{l,j+1})\big), \\
&c^{(k+1)}_{l,j+1}={\rm Pr}_{[-\nu_{l,j}, \nu_{l,j}]}\big(c^{(k)}_{l,j+1} - \alpha_k \, {\rm ApprGr}(c^{(k)}_{l,j+1}) + \beta_k \big(c^{(k)}_{l,j+1} - c^{(k-1)}_{l,j+1} \big) \big),\\
&c^{(k+1)}_{l,j+1}={\rm Pr}_{[-\nu_{l,j}, \nu_{l,j}]}\big(c^{(k)}_{l,j+1} - \alpha_k \, {\rm ApprGr}(c^{(k)}_{l,j+1}) + \beta_k \big(c^{(k)}_{l,j+1} - c^{(k-1)}_{l,j+1} \big)  +\\
&\quad\quad\quad\quad\quad\quad\quad\quad\quad\quad\quad\quad+\gamma_k(c^{(k-1)}_{l,j+1} - c^{(k-2)}_{l,j+1}) \big).  
\end{align*}
If it is needed, one can make a~good number of various initial controls.

The formula (Article~I,~4.1) is exact for any~$M$, but $M$ should be good for constructing a~sufficiently appropriate PConst. $\sigma$. Here (\ref{sect7_f12}), (\ref{sect7_f13}) are used for (\ref{sect7_f2}), (\ref{sect7_f7}) correspondingly. If $M$ is sufficiently large and $\Theta_M$ is uniform, then one can simply compute the infin.-dim. gradient~(Article~I,~3.3) at the midpoints and adjust the GPM parameters in the iterative GPM schemes (Article~I,~3.6)---(Article~I,~3.8) given for operating in the functional class of controls. If $M$ is significantly decreased, then it is needed to change (\ref{sect7_f12}), (\ref{sect7_f13}) to some composite quadrature formulas, generalize~Lemma~1. 

\vspace{-0.1cm}
\section{GPM, GA for the Transfer and Keeping Problems, $N=3$}
\label{Sect3}
\vspace{-0.1cm}

The given below examples show the GPM, GA results obtained by the author using the corresponding Python programs written by him using various needed standard tools from the libraries {\tt NumPy}, {\tt SciPy}, {\tt Matplotlib}, etc. (including, e.g., {\tt scipy.linalg.expm}), the GA implementation~\cite{Solgi_Genetic_algorithm_Python}. \vspace{0.1cm}

\begin{table}[h!!] 
\vspace{-0.2cm} 
\footnotesize
\centerline{Table~1. For Example~1.} 
\vspace{0.1cm}
\centering 
\begin{tabular}{ | c | l | c | c | } \hline
Case & GPM form & No.~of ${\bf a}^{(0)}$ & Complexity \\ \hline
1 & 2S, $\alpha=1, \beta=0.9$ & 1 & 3641 \\ \hline
2 & 1S, $\alpha=1$ & 1 & 44553~(!)\\ \hline
3 & 3S, $\alpha=1, \beta=0.9, \gamma=0.03$ & 1 & 2093 \\ \hline
4 & 2S, $\alpha=1, \beta=0.9$ & 2 & 6457 \\ \hline
5 & 1S, $\alpha=1$ & 2 & 63653~(!) \\ \hline
6 & 3S, $\alpha=1, \beta=0.9, \gamma=0.03$ & 2 & 4579 \\ \hline
\end{tabular}
\vspace{-0.1cm} 
\label{table1}
\end{table} 
\normalsize 
\begin{figure}[h!!]
\centering
\vspace{-0.1cm}
\includegraphics[width=1\linewidth]{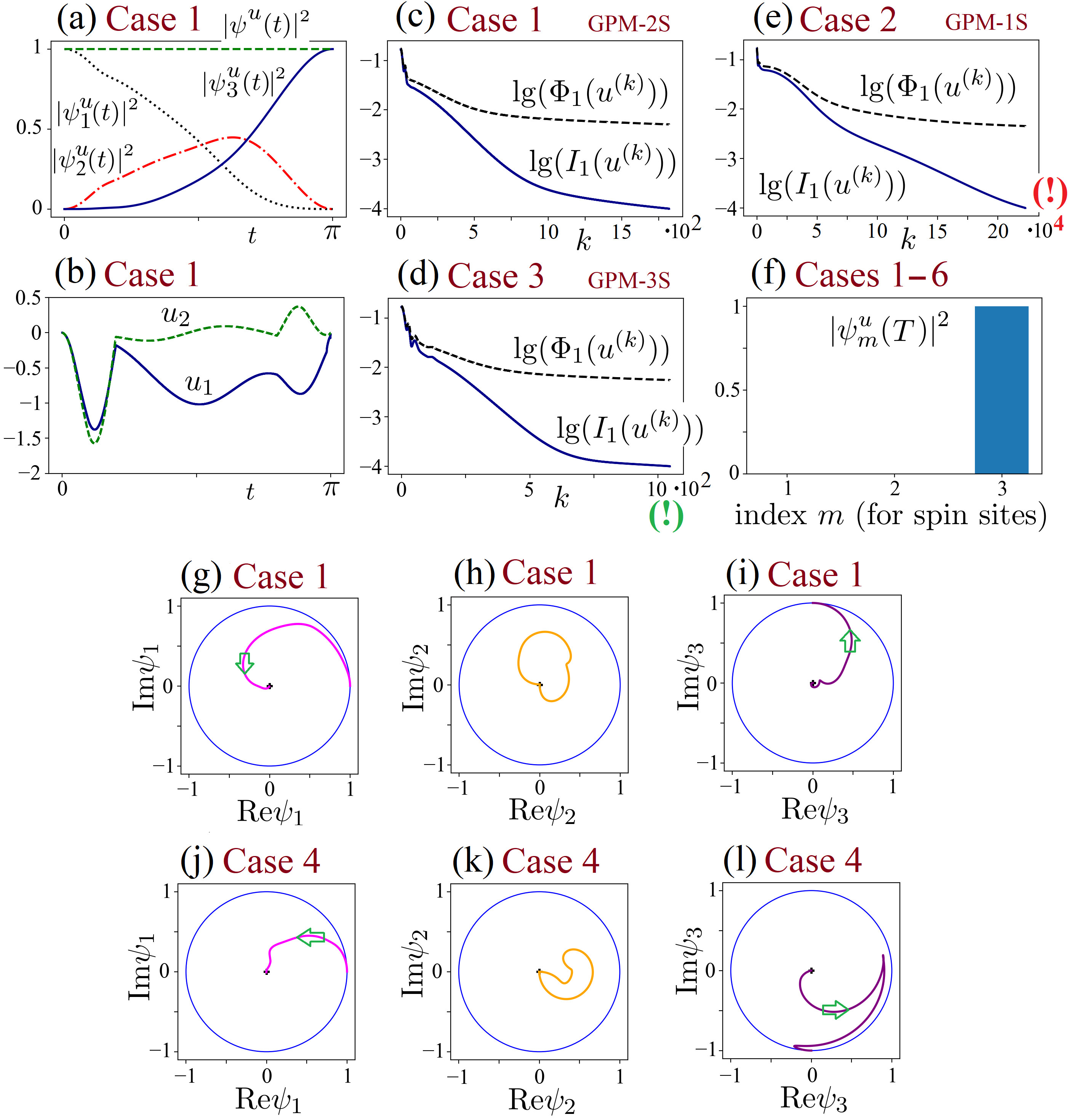}
\caption{For Example~\ref{example1}: (a),\,(b),\,(g)--(i) in Case~1, the provided by GPM-2S excitement's exchange in the terms of the resulting $\{\psi_m\}_{m=1}^3$ and controls; (c),\,(e),\,(d)~in Cases~1,~2,~3, the lg-values of $\Phi_1,~I_1$ at the GPM iterations; (j)--(l) in Case~4, the resulting $\{\psi_m\}_{m=1}^3$ obtained via GPM-2S; (f)~in Cases~1--6, the signal is (approximately) concentrated at the last spin site. \label{Fig1} }  \vspace{-0.2cm} 
\end{figure} 

~\par{\bf Example 1.} 
\label{example1}
Consider the transfer problem with $f_1({\bf a}) = \Phi_1(u)$ with $T = \pi$. Take $b_1, b_2$ with $\overline{b}_l = 5$, $q_l=8$, $l=1,2$ for (Article~I,~2.4). Take $P_{u_l} = 6.5 \cdot 10^{-5}$, $C_{S_l}=25$, $l=1,2$ in (Article~I,~2.9). Consider $M = 1570$ and PConst. $\sigma$ and controls with $\Delta t = T/M$. Consider the fin.-dim. GPM-1S(fixed $\alpha$), GPM-2S(fixed $\alpha,\beta$), GPM-3S(fixed $\alpha,\beta,\gamma$). Each iteration of such a~GPM form needs to solve the corresponding 2 Cauchy problems, just for the quantum and adjoint systems; these problems were solved via the formulas given in [Lemma~2, Article~I] and Lemma~1. We estimate the complexity of such a~GPM as the $2k+1$ solved Cauchy problems for $k \geq 0$. Table~1 shows the comparative GPM results under the two cases of ${\bf a}^{(0)}$. In Cases~1--3, we take ${\bf a}^{(0)}$ representing $u^{(0)}$ of the form~(Article~I,~2.6) with $\widehat{t}_1 = 0.1 T$, $\widehat{t}_2 = 0.2 T$, $\widehat{t}_3 = 0.8 T$, $\widehat{t}_4 = 0.9 T$, $\theta_1^{\rm L} = \theta_1^{\rm R} = -0.2$, $\theta_2^{\rm L} =-0.2=-\theta_2^{\rm R}$,  $y_1=-0.1$, $y_2=0$. In Cases~4--6, we take ${\bf a}^{(0)}$ obtained by approximating (at the same $\Theta_M$) $u^{(0)}$ being symmetrical --- relatively to the abscissa axis --- for $u^{(0)}$ taken in Cases~1--3. We see that, first, GPM-2S is faster --- {\it only via the Polyak's term} (i.e. with $\beta$) --- than GPM-1S by many times and, second, GPM-3S is essentially faster than GPM-2S. Also Fig.~1 shows the obtained results. GPM-3S is faster than GPM-1S in $\approx 21$ and $\approx 14$ times under the same $\alpha=1$ in Cases~2,~3 and in Cases~5,~6, correspondingly. We see the non-uniqueness of the problem's solving. 

~\par{\bf Example 2.} 
\label{example2}
Consider the problem to keep $\psi_0=\psi_{\rm g}$ at $[0,T=0.5]$. 

\vspace{-0.2cm}
\begin{figure}[h!]
\label{Fig2}
\centering
\includegraphics[width=0.7\linewidth]{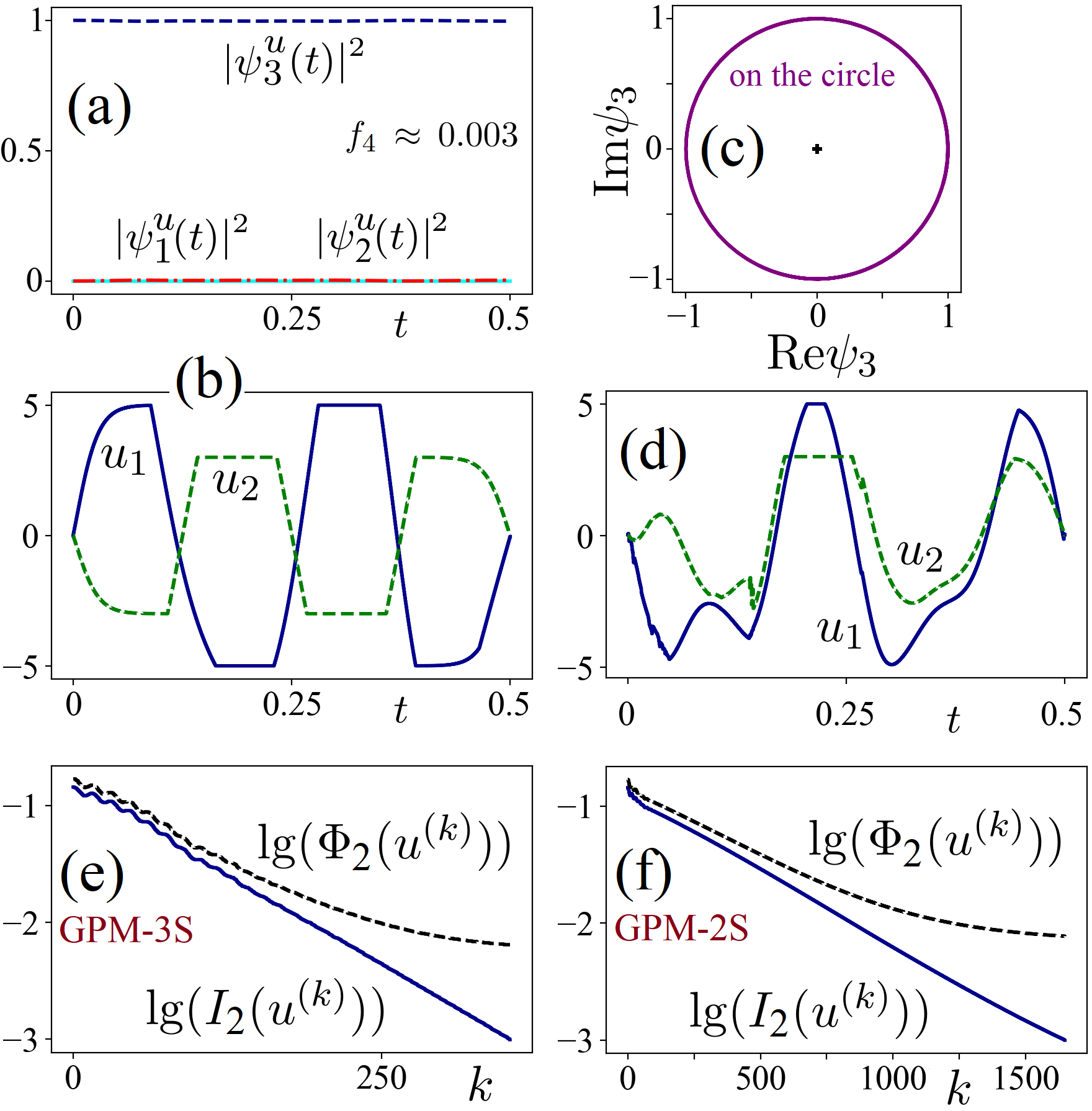}
\caption{For Example~2: (a)--(c)~the results obtained via GA for the approximated (as PConst.) class of controls~(2.11); (d),\,(e)~the resulting controls and the iterative behavior of GPM-3S; (f)~the iterative behavior of GPM-2S.}
\end{figure}  

{\it GA.} The constraint (Article~I,~2.4) is taken with $\overline{b}_1 =5$, $\overline{b}_1 =3$, $q=8$. Consider the class (Article~I,~2.11) with $M_{\sin}=3$ and $\gamma_{l,i} \in [-\overline{b}_l, \overline{b}_l]$, $\omega_{1,i} \in [3, 10]$, $\omega_{1,i} \in [2, 4]$ with $l=1,2$, $i=\overline{1,M_{\sin}}$ that form the set $Q_{\bf y} \ni {\bf y}=(\gamma_{1,1}, \omega_{1,1}, ..., \gamma_{2,M_{\sin}}, \omega_{2,M_{\sin}})$. In order to use the exact formula (Article~I,~4.1), the special class was approximated at the uniform grid $\Theta_M$ with $M=500$. It is needed to minimize $f_4({\bf y})$ introduced in~(Article~I,~2.12) where take $P_{\bf y}=0$. GA was used six times, independently on each other, with automatically generated ${\bf y}^{(0)}$, where the best result is $f_4 \approx 0.003$ which is {\it impressive} because the class is simple. The GA implementation \cite{Solgi_Genetic_algorithm_Python} was used; the GA parameters {\tt max\_num\_iteration} and {\tt population\_size} were taken correspondingly equal to 300 and~100. During the best trial GA run, the values 0.01, 0.005 of $\max\limits_{j \in \overline{1,M}}\{ F(\psi^{u^{(0)}}(t); \psi_{\rm g} \}$ were approximately reached at the cost correspondingly of the 195, 1430 solved Cauchy problems. Fig.~2(a)--(c) show the obtained best GA result. Subfig.~(a) shows that $|\psi_m^u(t)|^2$, $m=1,2,3$ are almost unchanged over~$[0,T]$. Subfig.~(b) shows the resulting controls. In~Subfig.~(c), we see that the obtained points $({\rm Re}\psi_3(t)$, ${\rm Im}\psi_3(t))$ with the noted precision are placed on the circle~$S^1(1)$. 

\begin{table}[h] 
\vspace{-0.1cm} 
\footnotesize
\begin{center} 
\centerline{Table~2. For Example~2.} 
\vspace{0.1cm} 
\begin{tabular}{ | l | c | c | c | c | c |} \hline
GPM & Complexity & $f_2(a)$ & $F(\psi^u(T); \psi_{\rm g})$ & $\int\nolimits_0^T F(...)dt$ & $\max\limits_{j \in \overline{1,M}}\{ F(...)\}$ \\ \hline
3S & 709 ($\approx$ 1.4\%) & $\approx 0.006$ & $\approx 0.001$ & $\approx 0.005$ & $\approx 0.022$ \\ \hline
2S & 3297 ($\approx$ 6.5\%) & $\approx 0.008$ & $\approx 0.001$ & $\approx 0.007$ & $\approx 0.029$ \\ \hline
1S & 50427 (100\,\%) & $\approx 0.008$ & $\approx 0.001$ & $\approx 0.007$ & $\approx 0.031$ \\ \hline
\end{tabular}
\end{center} \vspace{-0.1cm} 
\end{table} 
\normalsize 

{\it GPM.} In $\Phi_2(u)$, take $P_{\psi}=1$, $P_{u_1}=P_{u_2}=0$. Consider (Article~I,~2.11) with $M_{\sin} = 3$, $\gamma_{1,1} = -3$, $\gamma_{1,2} = -2$, $\gamma_{1,3} = 1$, $\omega_{1,1} = 4$, $\omega_{1,2} = 8$, $\omega_{1,3} = 5$, $\gamma_{2,1} = -4$, $\gamma_{2,2} = -3$, $\gamma_{2,3} = -2$, $\omega_{2,1} = 3$, $\omega_{2,2} = 4$, $\omega_{2,3} = 2$ and $b_1,b_2$ with $\overline{b}_1 =5$, $\overline{b}_1 =3$, $q=8$. Approximate these controls at the uniform grid $\Theta_M$ with $M=1000$. Thus, we have the initial point~${\bf a^{(0)}}$ so that $f_2({\bf a^{(0)}}) = \Phi(u^{(0)}) \approx 0.170$, $F(\psi^{u^{(0)}}(T); \psi_{\rm g}) \approx 0.144$, $\int\nolimits_0^T F(\psi^{u^{(0)}}(t); \psi_{\rm g})dt \approx 0.025$, and $\max\limits_{j \in \overline{1,M}}\{ F(\psi^{u^{(0)}}(t); \psi_{\rm g} \} \approx 0.144$. Consider GPM-$j$S, $j=1,2,3$ with the stopping condition $F(\psi^u(T);\psi_{\rm g}) < 10^{-3}$ $\&~\int\nolimits_0^T F(\psi^u(T);\psi_{\rm g})dt < 8 \cdot 10^{-3}$. In general, one can expect that the GPM results are essentially depend on adjusting $\overline{b}_1,\overline{b}_2$ in (Article~I,~2.4), $P_{\psi}$ in $\Phi_2$ and $\alpha_k, \beta_k, \gamma_k$ in~GPM. Consider GPM-3S($\alpha=2, \beta=0.93, \gamma=0.05)$, GPM-2S($\alpha=2, \beta=0.93)$, and GPM-1S($\alpha=2$). Table~2 compares the three GPM forms, here their complexity is estimated as how many the Cauchy problems were solved. Fig.~2(d)--(f) show the GPM-3S, GPM-2S iterations; these GPM forms show the large gains in comparison to GPM-1S.  

\vspace{-0.35cm}
\section{Non-Assigned~$T$ for a~Transfer. The Case $N=20$ and~GA}
\label{Sect4}
\vspace{-0.1cm}

Consider $T \in [T_1,T_2]$ and (via the known approach used, e.g., in \cite{Morzhin_AiT_2012} (O.V.~Morzhin, 2012)) change time $t \in [0,T]$ to $\tau \in [0,1]$ so that $t = T\,\tau$, $dt=T\,d\tau$. Consider $v=(\overset{\smallsmile}{u},T)$, $\overset{\smallsmile}{u}(\tau)=u(T\tau)$, $\overset{\smallsmile}{\sigma}(\tau)=\sigma(T\tau)$,~etc., and \vspace{-0.46cm}
\begin{equation}
\frac{d\overset{\smallsmile}{\psi}\,^v(\tau)}{d\tau} = -i\,T\,\big(H_0 + \overset{\smallsmile}{H}_1(\tau,\overset{\smallsmile}{u}(\tau)) \overset{\smallsmile}{\psi}\,^v(\tau),~~ \tau \in [0,1],~~\overset{\smallsmile}{\psi}\,^v(0)=\psi_0
\label{sect9_f1}
\vspace{-0.2cm}
\end{equation} 
for an arbitrary~$N$. Consider the corresponding change in the special class~(Article~I,~2.6), approximate it with a~grid $\{\tau_0=0 < ... < \tau_j < ... < \tau_M=1 \}$, and then, instead of $f_3({\bf x})$ used in (Article~I,~2.10) for the transfer requirement, take \vspace{-0.4cm}
\begin{align}
\overset{\smallsmile}{f}_3({\bf x},T) &= 1- \big|\langle \overset{\smallsmile}{\psi}\,^v(\tau=1), \psi_{\rm g} \rangle \big|^2 + P_{\bf x} \sum\nolimits_{j=1}^{M-1} \sum\nolimits_{l=1}^2 |\overset{\smallsmile}{u}_l(\tau_j; {\bf x},T)| + \nonumber \\
& \quad  + P_T\,T \to \min\limits_{({\bf x},T)}, ~~ P_{\bf x},~P_T \geq 0.
\label{sect9_f2} 
\end{align}

\vspace{-0.2cm}

~\par{\bf Example 3.} 
\label{example3}
For the case $N=20$, consider the transformed system (\ref{sect9_f1}), which contains $T$ as a~controlling parameter, and the problem~(\ref{sect9_f2}), uniform grid $\{\tau_j\}_{j=1}^M$ with $M=500$ at $[0,1]$. The corresponding program written by the author in Python uses the GA implementation~\cite{Solgi_Genetic_algorithm_Python}. For example, take $T \in [23,26]$, $P_T=0$ (i.e. simply free $T$), $P_{\bf x}=0$, etc. Fig.~3 shows the results of some one among several trial GA runs. Although the class of controls is simple, the final transfer infidelity is $1- \big|\langle \overset{\smallsmile}{\psi}\,^v(\tau=1), \psi_{\rm g} \rangle \big|^2 \approx 0.009$, i.e. only $\lessapprox 1$~\% of the infidelity $1- \big|\langle \overset{\smallsmile}{\psi}\,^v(\tau=0), \psi_{\rm g} \rangle \big|^2 =1$; the corresponding $T \approx 24.9$. The GA parameters {\tt max\_num\_iteration} and {\tt population\_size} were taken correspondingly equal to 400 and~50. 
\begin{figure}[h!!]
\centering
\vspace{-0.2cm}
\includegraphics[width=1\linewidth]{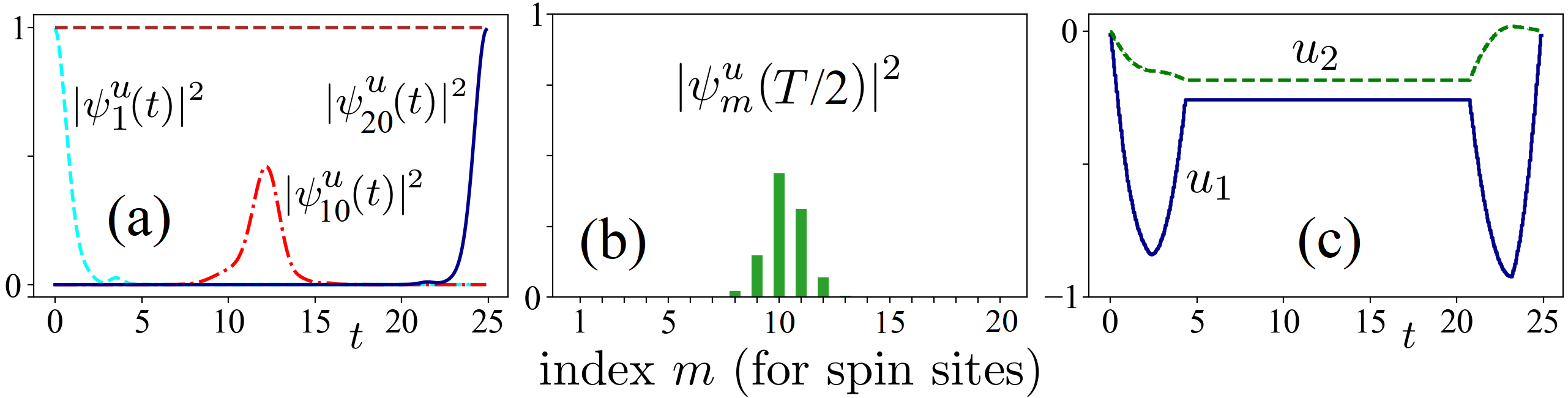}
\caption{For Example~3 with $N=20$, the GA results  (before adding normally distributed noises): (a),\,(b) transfer in the terms of $|\psi_m^u(t)|^2$; (c)~resulting~controls. \label{Fig5} } 
\end{figure} 

\vspace{-0.2cm}
\begin{table}[h]  
\footnotesize
\begin{center} 
\centerline{Table~3. For Example~3. The statistical results over a lot of cases of noised controls.} 
\vspace{0.1cm} 
\begin{tabular}{ | l | c | c | c | c | c | c |} \hline
$\sigma$ in $\mathcal{N}$ & $\min Y$ & $\max Y$ & $\min\,W$ & $\max\,W$ & ${\rm mean}\,W$ & ${\rm median}\,W$ \\ \hline
0.05 & $\approx-0.26$ & $\approx0.27$ & $\approx 0.008$ & $\approx 0.036$ & $\approx 0.014$ & $\approx 0.014$ \\ \hline
0.1 & $\approx-0.52$ & $\approx0.50$ & $\approx 0.008$ & $\approx 0.113$ & $\approx 0.032$ & $\approx 0.029$  \\ \hline
0.15 & $\approx-0.77$ & $\approx0.79$ & $\approx 0.013$ & $\approx 0.212$ & $\approx 0.061$ & $\approx 0.056$  \\ \hline
0.2 & $\approx-1.1$ & $\approx1.03$ & $\approx 0.017$ & $\approx 0.407$ & $\approx 0.101$ & $\approx 0.093$ \\ \hline
\end{tabular}
\end{center} 
\end{table} 
\normalsize 
\vspace{-0.2cm}

After that the following analysis was performed. In each $p$th run among $10^4$ independent runs, random values $\{n_{l,j,p}\}_{j=0}^{M-1}$ ($M=500$, $l=1,2$) were generated with the normal distribution $\mathcal{N}(\sigma=0.05,\mu=0)$ and were added to the PConst. $u_1,\,u_2$ obtained above via GA, i.e. we have $2 M\cdot 10^4$ values $\{u_l(\tau_j) + n_{l,j,p}\}$ of $2 \cdot 10^4$ such {\it noised} controls and compute the corresponding $10^4$ values $W=\big\{1- \big|\langle \overset{\smallsmile}{\psi}\,^v(1), \psi_{\rm g} \rangle \big|^2\big\}$. The size of the corresponding $Y = \{n_{l,j,p}\}$ is $2M \cdot 10^4$. The value $T$ is the same. Such the computations were performed also for $\sigma=0.1,\,0.15,\,0.2$ in $\mathcal{N}$. Thus, $4 \cdot 10^4$ times the quantum equation was solved using~[Lemma~2, Article~I] and the $2 \cdot 10^7$ matrix exponentials ($20 \times 20$) were computed here. Table~3 shows the obtained min, max of $Y$ (tailed values) and min, max, arithmetic mean, median of $W$ for each $\sigma \in \{0.05, 0.1, 0.15, 0.2\}$. We see that some mild noise does not much change the final infidelity obtained via the GA~approach.  

\vspace{-0.1cm}
\section{Some Methodological Comments}
\label{Sect5}
\vspace{-0.4cm}

~\par {\bf With respect to the pointwise constraint (Article~I,~2.4).} 
Some part of papers in QOC does not restrict controls' values from below and above, e.g., the paper~\cite{Michoel_Mulherkar_Nachtergaele_NJP_2010} which considers the Schr\"{o}dinger equation with linearly entering controls and is about generating quantum gates in the terms of an another type of spin chains and uses a~gradient descent (p.\,9) with a~small step size. In a good number of papers, controls are pointwisely constrained, including, e.g., \cite{Pechen_Rabitz_2006}. This article considers (Article~I,~2.4) (by analogy with \cite{Morzhin_Pechen_JPA_2025}) and (Article~I,~2.5) and takes them exactly into account in the GPM forms (Article~I,~3.6)--(Article~I,~3.8) and (\ref{sect7_f9})--(\ref{sect7_f11}), correspondingly.  

\vspace{0.2cm}

{\bf Extremal control.} If a non-optimal control $u^{(0)}$ satisfies (Article~I,~3.5), then GPM-1S starting from $u^{(0)}$ cannot improve $u^{(0)}$. Because GPM-2S starts from the result $u^{(1)}$ of the initial iteration of GPM-1S, then GPM-2S using $u^{(1)}(t) - u^{(0)}(t) \equiv 0$ cannot make less than $\Phi_p(u^{(0)})$ in this way.

\vspace{0.2cm}

{\bf Polyak's heavy-ball method.} This method is long known \cite{Polyak_ZVMMF_1964} (B.T.~Polyak, 1964) in the mathematical theory of unconstrained optimization and is an~{\it origin} for the considered GPM-2S, GPM-3S. In this regard, consider the following quote which is taken from \cite{Polyak_book_1983_1987} (B.T.~Polyak, 1987) and is about minimizing a~differentiable function $f(x)$. ``Clearly, if $\nabla f(x^k) = 0$, but $x^k \neq x^{k-1}$, then $x^{k+1} \neq x^k$, i.e., the method does not jam at a stationary point. By the mechanical analogy ... the motion of a heavy ball along the uneven surface, it follows that if the velocity of the ball is sufficiently large, it ``skips'' over shallow holes. It is possible to show by examples that this method does indeed have the property that it escapes the shallow local minima. However, it may ``fall'' into a deep minimum and would not be able to get out.'' About this method, consider also the following quote from~\cite[p.~351]{Strang_book_2019} (G.~Strang, 2019): ``So we add momentum with coefficient~$\beta$ to the gradient (Polyak's important~idea)''. The next quote is from \cite{Mordukhovich_Nemirovski_Nesterov_2026} (B.S.~Mordukhovich, A.S.~Nemirovski, Yu.E.~Nesterov, 2026): ``...\,Heavy Ball Method ... has become extremely useful and popular in modern research and applications in the areas of machine learning and artificial intelligence~...'' The recent work \cite{Lopatin_Kozyrev_Pechen_2025}~(I.A.~Lopatin, S.V.~Kozyrev, A.N.~Pechen, 2025) also considers such an optimization which is able to make an improvement  under a~gradient equal to~zero. 

\vspace{0.2cm}

{\bf Without global assuming convexity / strong convexity.} The current article does not assume convexity / strong convexity at the whole convex~$Q_{\bf a}$ for the considered objective functions $f_p({\bf a})$, $p=1,2$. In the theory of constrained fin.-dim. optimization, note, e.g., the following. Theorems~5,~6 in \cite{Antipin_DifferEqu_1994}\,(A.S.~Antipin, 1994) for the fin.-dim. {\it 2nd order GPM}, --- i.e. for the continuous analog of the fin.-dim. GPM-2S given in \cite[Eq.~(3.3)]{Antipin_DifferEqu_1994}, --- assume that a~smooth objective function $f(x)$ with the Lipschitz continuous gradient is {\it convex or strongly convex} (the set~$Q$ is convex). In \cite{Nedich_1993}\,(A.~Nedich, 1993), Theorems~1,\,2 for GPM-3S and Theorems~3,\,4 for GPM-2S also assume that an~objective function is {\it convex or strongly convex}. In \cite{Nedich_DiffEq_1994}\,(A.~Nedich, 1994), Theorems~1,~2 for the given there fin.-dim. {\it third-order GPM} assume that an~objective function is {\it convex or strongly convex},~etc. In nonconvex fin.-dim. optimization, \cite{Zavriev_Kostyuk_1993}~(S.K.~Zavriev, F.V.~Kostyuk,~1993) considers the heavy-ball method in the terms of the Lyapunov direct~method. Nowadays, studying local traps of various orders for controls landscapes forms an important direction in QOC~\cite{Pechen_Tannor_PRL_2011}~(A.N.~Pechen, D.J.~Tannor, 2011), \cite{Pechen_LJM_2025}~(A.N.~Pechen,~2025). 

\vspace{-0.1cm}
\section{Conclusion-II}
\label{Sect6}
\vspace{-0.4cm} 

~\par This article (II) continues Article~I. Based on the infin.-dim. gradients given in [Article~I, Lemma~1], the suggested in Lemma~1 fin.-dim. gradients were derived. Theorem~1 adapts some corresponding projection-type necessary condition for optimality. The corresponding GPM forms are given in Sect.~\ref{Sect2}. The obtained (by the author) numerical results described in Articles~I,~II for the transfer and keeping models show the usefulness of the adapted GPM, GA for these models and increase the interest to continue this research. In particular, Article~II shows that GPM-2S and GPM-3S can be significantly faster in comparison to GPM-1S. Sect.~\ref{Sect5} gives some important methodological comments. The results of Articles~I,~II were briefly described in the conference's talk \cite{Morzhin_Talk_Nov28_2025}~(O.V.~Morzhin,~2025).

{\small {\bf Acknowledgment.} The author is grateful to A.N.~Pechen (Moscow), Dr.~Phys.-Math. Sci., RAS Prof., for a good discussion of this article.}

\end{document}